\documentclass[twocolumn]{jpsj2}

\def\runtitle{Superconductivity in Co-oxides}
\def\runauthor{Yasuhiro \textsc{Tanaka}, Youichi \textsc{Yanase} 
and Masao \textsc{Ogata}}

\title{%
Superconductivity in Na$_x$CoO$_2\cdot y$H$_2$O 
by charge fluctuation
}

\author{%
Yasuhiro \textsc{Tanaka},\thanks{yasuhiro@hosi.phys.s.u-tokyo.ac.jp}
Youichi \textsc{Yanase} 
and Masao \textsc{Ogata}\thanks{ogata@phys.s.u-tokyo.ac.jp}
}

\inst{%
Department of Physics, University of Tokyo, 
Hongo, Bunkyo-ku, Tokyo 113-0033 Japan
}

\recdate{November 12, 2003}

\abst{%
A new mechanism for superconductivity in the newly discovered 
Co-based oxide is proposed by using charge fluctuation.
A single-band extended Hubbard model on the triangular lattice 
is studied within the random phase approximation.  
$f$-wave triplet superconductivity is stabilized in the vicinity 
of charge-density-wave instability, which is in sharp contrast with 
the square-lattice case.  
The physical origin of the realization of the $f$-wave triplet state 
as well as the relevance to experiments are discussed.  
}
\kword{%
Co-based oxides, charge fluctuation, triplet superconductivity
}

\begin{document}
\sloppy
\maketitle

Recently discovered superconductivity in 
Na$_x$CoO$_2$ $\cdot y$H$_2$O\cite{Takada,Cava} 
has attracted great interest since it is realized in 
a frustrating system.
Compared with high-$T_c$ cuprates and Ru-based superconductivity, 
Sr$_2$RuO$_4$, Co oxides have various unique and interesting
properties. 
Firstly this is the first Co-based oxide superconductivity.  
Secondly Co atoms form two-dimensional triangular lattice, which 
is in sharp contrast with the high-$T_c$ cuprates or Sr$_2$RuO$_4$.  
Thirdly there are three bands close to the Fermi surface.
Recent intensive experiments have shown that the superconductivity 
is unconventional\cite{Waki,Nagoya,Fujimoto,Ishida} 
and the material has strong correlation.\cite{Ong,Yang}

Superconductivity in the systems with frustration is a 
very interesting subject.  
Originally the resonating valence bond (RVB) idea by Anderson was 
proposed in the frustrating triangular Heisenberg spin system.\cite{PWA}  
This concept was developed to the idea of RVB superconductivity 
which is expected to be realized in the doped Heisenberg system 
like high-$T_c$ superconductivity.\cite{PWA2}  
Since the high-$T_c$ cuprates are based on the square lattice 
without frustration, superconducting materials 
in frustrating systems have been long desired.  
The new superconductor, Na$_x$CoO$_2$$\cdot y$H$_2$O, is in 
this sense an ideal material which has a perfect triangular 
lattice of Co atoms.  
Basically frustration destroys the long-range order 
like antiferromagnetism and induces large fluctuations.  
There is a potential possibility that there appears high-$T_c$ 
superconductivity using these large fluctuations.  
One candidate in the Co oxides is spin fluctuation, 
which has been discussed using $t$-$J$ models on the triangular 
lattice.\cite{Hu,Baskaran,Shastry,Lee,Ogata}  
Another candidate is charge fluctuation 
which is realized in the extended Hubbard model.
They contain new physics induced by frustration.  
In the present paper, we focus on the latter possibility in 
Na$_x$CoO$_2$$\cdot y$H$_2$O system.  

Let us first discuss the effects of multi-orbitals.\cite{Singh,Koshibae}  
In the case of Sr$_2$RuO$_4$, the three orbitals 
(d$_{xy}$, d$_{yz}$ and d$_{zx}$) do not mix so much.  
On the contrary, in the present material, Na$_x$CoO$_2$, 
the three orbitals mix relatively strongly and three bands 
are formed near the Fermi energy.  
However it has been pointed out that the main band centered 
around the $\Gamma$ point ($\mib{k}=(0,0)$) is made out of the 
symmetric combination of the orbitals, i.e., $a_{2g}$-orbital, or
$\frac{1}{\sqrt{3}}({\rm d}_{xy}+{\rm d}_{yz}+{\rm d}_{zx})$.\cite{Singh}  
Thus, in the present paper, we use a single band effective 
Hamiltonian as a first approximation.  

The $t$-$J$ model on the triangular lattice has been extensively 
studied by various groups.\cite{Hu,Baskaran,Shastry,Lee,Ogata}  
However every mean-field-type theory has concluded that 
the most probable RVB superconductivity has 
$d_{x^2-y^2}+$i$d_{xy}$-wave symmetry.  
This state has been confirmed by high-temperature expansion 
studies.\cite{Koretsune2} 
However, recent $\mu$SR experiment\cite{Higemoto} 
has shown that no evidence for the broken time-reversal-symmetry 
is found, which contradicts with the prediction in the $t$-$J$ model.  
Alternatively it has been proposed that charge ordering is 
realized near the electron density $n=1+1/3$ and $n=1+2/3$
(more-than-half-filling).\cite{Baskaran2}  
It is apparent that the 1/3- or 2/3-filling is a special filling 
for the triangular lattice.  
Motrunich and Lee\cite{Motrunich} 
have discussed a strongly correlated electron 
system just next to the charge ordering.  


Having these in mind, we discuss superconductivity 
due to charge fluctuation in the vicinity 
of charge density wave (CDW).  
We use a single-band extended Hubbard model on the triangular lattice 
which is given by
\begin{equation}
\label{eq:Hamiltonian}
H = \sum_{\mib{k} \sigma} \varepsilon_{\mib{k}} c_{\mib{k} \sigma}^{\dagger}
c_{\mib{k} \sigma} +U \sum_i n_{i\uparrow} n_{i\downarrow} 
 +  V \sum_{\langle ij\rangle} n_i n_j \ ,
\end{equation}
where $c_{\mib{k} \sigma}^{\dagger}$ and $c_{\mib{k} \sigma}$ denote 
electron creation and annihilation operators respectively and 
$\langle ij\rangle$ represents the 
summation over the nearest neighbor pairs. 
$\varepsilon_{\mib{k}}$ is the 
dispersion relation of the triangular lattice, namely
$\varepsilon_{\mib{k}} = 
-2t(\cos k_y + 2\cos \frac{\sqrt{3}}{2}k_x \cos \frac{1}{2}k_y)-\mu$.
In this extended Hubbard model, we expect a CDW state when $V$ is 
very large and $U>V$.  
For the intermediate $V$, superconductivity 
is predicted in the vicinity of the CDW 
instability.\cite{Scalapino,McKenzie,Kobayashi}
Since previous theories only discussed the square or cubic lattice, 
it is necessary to study the case of triangular lattice.  
Surprisingly we find that the triplet $f$-wave state is stabilized, 
instead of $d_{xy}$-wave obtained in the square lattice.

The LDA calculation\cite{Singh} showed that there is a large hole-like 
Fermi surface around the $\Gamma$ point so that the effective 
hopping integral is $t<0$.  
For the Co oxides, the electron density $n$ is more than unity 
($n$ is around $1.3$).  
In the following, we use a convention with $t>0$ and $n<1$, 
which is equivalent to the case with $t<0$ and $n>1$ by 
electron-hole transformation, $c_{\mib{k}} \rightarrow c_{\mib{k}}^\dagger$.
The ``hole-doped case'' in the present paper 
corresponds to the Co oxides.  
We also study the ``electron-doped case'' for comparison.  

To study the effective interaction between electrons arising from
charge and spin fluctuations, we use random phase 
approximation (RPA).\cite{Scalapino,McKenzie,Kobayashi}
In RPA, the effective pairing potentials for 
the singlet and triplet channels are
\begin{align}
    \label{eq:s_pairing}
V^s(\mib{q}, \omega_l ) 
&= U +V(\mib{q}) + \frac{3}{2}U^2 \chi_s (\mib{q}, \omega_l ) 
               \nonumber \\
&- \bigl\{\frac{1}{2}U^2 +2U V(\mib{q})
              +2V(\mib{q})^2 \bigr\} \chi_c (\mib{q}, \omega_l),
\end{align}
\begin{align}
    \label{eq:t_pairing}
V^t(\mib{q}, \omega_l ) 
&= V(\mib{q}) -\frac{1}{2}U^2 \chi_s (\mib{q}, \omega_l ) 
               \nonumber \\
&- \bigl\{\frac{1}{2}U^2 +2U V(\mib{q})
              +2V(\mib{q})^2 \bigr\} \chi_c (\mib{q}, \omega_l),
\end{align}
where $V(\mib{q})=
2V(\cos q_y + 2\cos \frac{\sqrt{3}}{2}q_x \cos \frac{1}{2}q_y)$ and
$\omega_l$ is the Matsubara frequency. 
$\chi_s$ and $\chi_c$ are spin and charge susceptibilities,
respectively.
These quantities are calculated within RPA as follows,
\begin{equation}
 \label{eq:chi_s}
\chi_s (\mib{q}, \omega_l ) = 
 \chi_0 (\mib{q}, \omega_l )/\bigl[1-U\chi_0 (\mib{q}, \omega_l )\bigr],
\end{equation}
\begin{equation}
  \label{eq:chi_c}
 \chi_c (\mib{q}, \omega_l ) = 
 \chi_0 (\mib{q}, \omega_l )
   /\bigl[1+\{U+2V(\mib{q})\} \chi_0 (\mib{q}, \omega_l )\bigr].
\end{equation}
Here $\chi_0$ is the bare susceptibility given by
\begin{equation}
 \label{eq:chi0}
\chi_0 (\mib{q}, \omega_l ) =
 \frac{1}{N}\sum_{\mib{p}}
\frac{f(\varepsilon_{\mib{p}+\mib{q}})-f(\varepsilon_{\mib{p}})}
{\omega_l-(\varepsilon_{\mib{p}+\mib{q}}-\varepsilon_{\mib{p}})} \ .
\end{equation}
Note that the terms proportional to $\chi_c$ in eqs.\ (2) and (3) 
represent effective pairing potentials due to charge fluctuation.  
This charge fluctuation contributes equally to $V^s$ and $V^t$ 
since it comes from charge degrees of freedom.

Figure 1 shows the momentum dependences of the spin and charge 
susceptibilities at $T=0.01$.  
In the following we use $t$ as a unit of energy and $k_{\rm B}=1$.  
We calculate both hole-doped ($n<1$) and 
electron-doped ($n>1$) cases, since there is no particle-hole symmetry in 
the triangular lattice. 
Let us discuss the hole-doped case with $n=0.8$ (Fig.\ 1(a)), which 
corresponds to the Co oxides.  
When $V=0$ (i.e., $(U,V)=(3.64,0)$), the spin 
fluctuation is much larger than the charge fluctuation because 
the system is close to the spin density wave (SDW) instability. 
$\chi_s$ has a peak at $\mib{Q}=(0,\frac{4}{3}\pi)=$(K-point) 
due to the nesting condition of the system.  
On the other hand, when $V\ne 0$, i.e., $(U,V)=(3.21,1.2)$, 
$\chi_c$ has a peak structure around $\mib{q}=\mib{Q}$ where 
$V(\mib{q})$ has its maximum value. 
This shows that $V$ induces the charge 
density wave (CDW) instability for large-$V$.

For the electron-doped case $(n=1.2)$, in contrast, the peak in 
$\chi_s$ for $V=0$ moves away from $\mib{Q}$ as shown in Fig.\ 1(b).  
Near the CDW instability i.e., $(U,V)=(2.0, 1.06)$, on the other
hand, $\chi_c$ becomes large at $\mib{q}=\mib{Q}$.  
Apparently the peaks of $\chi_s$ and $\chi_c$ are located
at different positions from each other in contrast to the 
hole-doped case where both spin and charge fluctuations become
large near $\mib{q}=\mib{Q}$.  
\begin{figure}
\begin{center}
\includegraphics[width=6cm]{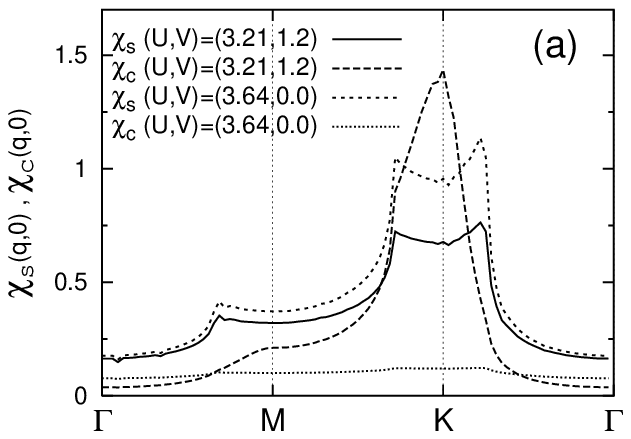}
\includegraphics[width=6cm]{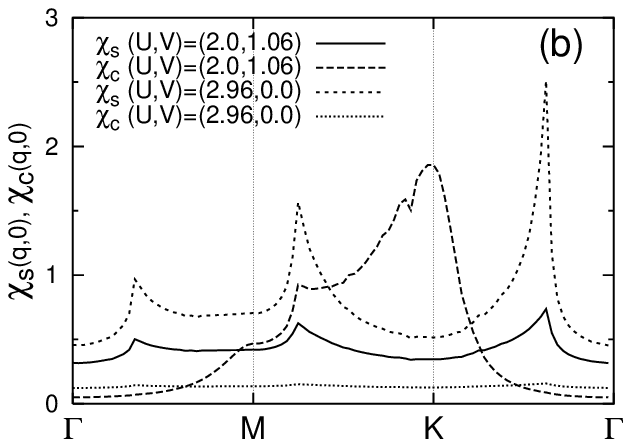}
\end{center}
\caption[]{
Spin ($\chi_s$) and charge ($\chi_c$) susceptibilities 
for (a) the hole doped $(n=0.8)$ and (b) the electron doped $(n=1.2)$
 cases at $T=0.01$. The parameters are chosen at $(U,V)=(3.64,0)$ and
$(3.21,1.2)$ in (a) and $(U,V)=(2.96,0)$ and $(2.0,1.06)$ in (b), 
respectively.}
\end{figure} 

Next, we calculate the pairing potentials, $V^s$ and $V^t$, 
as shown in Fig.\ 2. 
For $V=0$, the momentum dependence of $V^s$ is similar to that of 
$\chi_s$ in both hole- and electron-doped cases because contributions 
from $\chi_c$ is negligible.  
$V^t$ has the same property as $V^s$ although the sign is
opposite and the magnitude is smaller than $V^s$.
Near the CDW instability ($V\ne 0$), on the contrary, 
both $V^s$ and $V^t$ show
a negative (i.e., attractive) peak at $\mib{q}=\mib{Q}$.
Note that constant shift in $V^s$ or $V^t$ does not affect 
anisotropic superconductivity.  
In the region away from $\mib{Q}$, $V^s$ remains positive due to
the contribution from the spin fluctuation, whereas $V^t$ is negative 
for almost all region because both spin and charge fluctuations
lead to the attractive interaction as shown in eq.\ (\ref{eq:t_pairing}).
\begin{figure}
\begin{center}
\includegraphics[width=6cm]{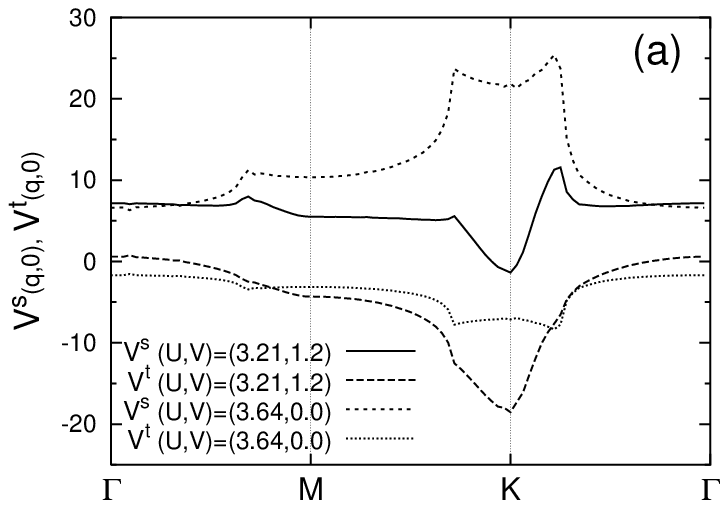}
\includegraphics[width=6cm]{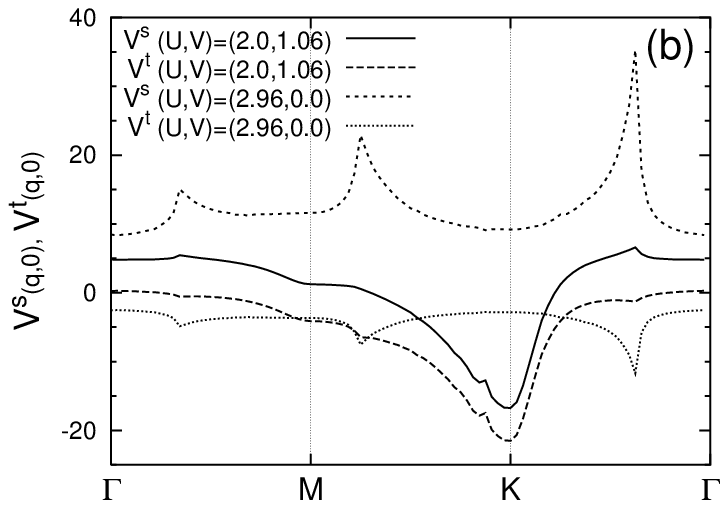}
\end{center}
\caption[]{
Effective interactions for singlet ($V^s$) and triplet ($V^t$) pairing 
for (a) hole-doped and (b) electron-doped cases, respectively 
at $T=0.01$. }
\end{figure}

To obtain the onset of the superconducting state, we solve the
linearized $\acute{\rm E}$liashberg's equation within the 
weak-coupling theory:
\begin{equation}
 \label{eq:Eliashberg}
 \lambda\Delta(\mib{k})=-\sum_{\mib{k}^{\prime}} 
V^{(s,t)}(\mib{k}-\mib{k}^{\prime},0)
 \frac{\tanh(\beta\xi_{\mib{k}^{\prime}}/2)}{2\xi_{\mib{k}^{\prime}}}
 \Delta(\mib{k}^{\prime}) \ ,
\end{equation}
with $\xi_{\mib{k}} = \varepsilon_{\mib{k}}-\mu$.
The transition temperature $T_c$ is determined by the condition, $\lambda=1$.
In the weak-coupling theory, $\omega$ dependence of the order parameter 
$\Delta(\mib{k})$ is neglected. Although this approximation is 
quantitatively insufficient, it is expected to be valid 
for investigating the pairing symmetry of 
$\Delta(\mib{k})$ and for grasping the basic idea of the superconductivity 
mediated by charge fluctuation.

The point symmetry group of the triangular lattice is D$_6$, 
which has six irreducible representations as shown in Table I. 
According to these representations, we classify $\Delta(\mib{k})$ or
eigenfunctions of $\acute{\rm E}$liashberg's equation.  
$\Delta(\mib{k})$ belonging to $A_1$, $A_2$ and $E_2$ symmetry 
correspond to spin-singlet states, whereas those belonging to 
$B_1$, $B_2$ and $E_1$ correspond to spin-triplet states. 
We find that $\Delta(\mib{k})$ with maximum eigenvalue belongs 
to $E_2$ for the spin-singlet case and 
$B_2$ for the spin-triplet case, respectively. 
This means that the realized superconductivity is $d$-wave singlet 
or $f$-wave triplet.  
Actually there are two degenerate states in the $d$-wave singlet channel
because it is two-dimensional representation ($E_2$).
These two states form d$_{x^2-y^2} +$id$_{xy}$ state below $T_c$ predicted 
in the $t$-$J$ model.  Interestingly we did not find $p$-wave state 
for the triplet channel in contrast to the Sr$_2$RuO$_4$ case where 
$p_x$+i$p_y$ state is realized.  

\begin{table}
\begin{center}
\caption{Irreducible representations (IR) of $D_6$}
\begin{tabular*}{80mm}{@{\extracolsep{\fill}}lr} \cline{1-2}
IR (symmetry)   &Basis functions\\ \cline{1-2}
$A_1$($s$) & 1 \\ \cline{1-2}
$A_2$($i$) & $\sin \frac{3}{2}\sqrt{3}k_x \sin \frac{k_y}{2} 
+\sin \frac{\sqrt{3}}{2}k_x \sin \frac{5}{2}k_y $
\\
\ & 
$ 
-\sin \sqrt{3}k_x \sin 2k_y$  \\ \cline{1-2}
$B_1$($f$) &  $\sin (\frac{k_y}{2})(\cos \frac{\sqrt{3}}{2} k_x
-\cos \frac{k_y}{2})$\\ \cline{1-2}
$B_2$($f$) &  $\sin (\frac{\sqrt{3}}{2} k_x) (\cos \frac{\sqrt{3}}{2} k_x
-\cos \frac{3}{2} k_y)$\\ \cline{1-2}
$E_1$($p$) & $\sin \frac{\sqrt{3}}{2} k_x \cos \frac{k_y}{2}$\\ \cline{1-2}
$E_2$($d$) & $\sin \frac{\sqrt{3}}{2} k_x \sin \frac{1}{2} k_y$\\ \cline{1-2}
\end{tabular*}
\end{center}
\end{table}
The obtained phase diagrams on the $(U,V)$ plane for $n=0.8$ and 
$n=1.2$ cases are shown in Figs.\ 3(a) and 3(b), respectively. The
temperature is fixed at $T=0.01$. The dashed lines which
correspond to the divergence of $\chi_s$ or $\chi_c$ determine
the boundary of SDW or CDW state, respectively. 

Let us discuss the hole doped case $(n=0.8)$ first. In Fig.\ 3(a), 
the eigenvalue with the $E_2(d)$ symmetry becomes larger than unity 
in the right-hand side of the solid line.  
In this case, the effect of $V$ suppresses the $d$-wave pairing,
which is understood as follows.  
For $n=0.8$ and without $V$, the effective interaction for 
the singlet pairing is large at $\mib{q}=\mib{Q}$ as shown in Fig.\ 2(a), 
inducing the $d$-wave superconductivity near the SDW.
In the presence of $V$, however, the charge fluctuation gives an 
opposite contribution at $\mib{q}=\mib{Q}$ and thus suppresses the pairing 
potential $V^s$, which results in the suppression of $d$-wave pairing.

\begin{figure}
\begin{center}
\includegraphics[width=6cm]{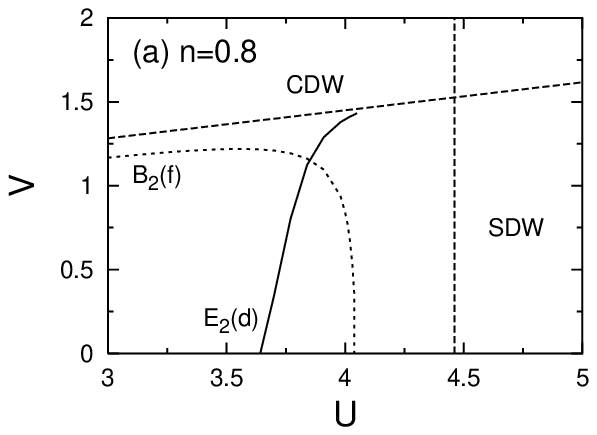}
\includegraphics[width=6cm]{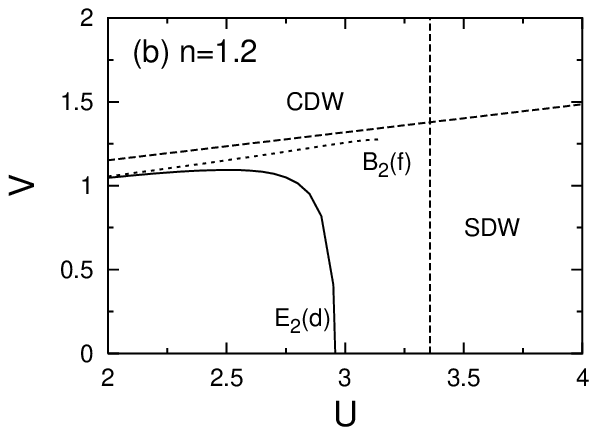}
\end{center}
\caption[]{
Phase diagram on the $(U,V)$ plane at $T=0.01$ for (a) hole-doped
and (b) electron-doped cases. The dashed lines correspond 
to the CDW and SDW instabilities.  
On the solid and the dotted lines, the eigenvalues which belong to 
E$_2$($d$) and B$_2$($f$) symmetry reach unity, respectively.
In (a), the $d$-wave pairing is dominant near SDW and the $f$-wave
pairing is dominant near CDW.  In (b), 
the eigenvalue of the $f$-wave solution reaches unity only in the
region where that of the $d$-wave solution is larger than unity.}
\end{figure} 

In the proximity to the CDW boundary, the superconductivity 
with $B_2(f)$ symmetry becomes dominant. The momentum dependence 
of the order parameter is shown in Fig.\ 4. The $f$-wave 
solution has three peaks with the same sign which are connected 
by the wave vectors $(0,\frac{4}{3} \pi)$, 
$(\frac{2}{\sqrt{3}} \pi, -\frac{2}{3} \pi)$ and $(\frac{2}{\sqrt{3}} 
\pi,\frac{2}{3} \pi)$ in the triangular lattice Brillouin zone. 
The order parameter with this property is favorable because of 
the momentum dependence of $V^t$ which gives large attractive 
interactions at these wave vectors. In fact, for 
the triplet pairing, both spin and charge fluctuations give
attractive interactions and work cooperatively since their magnitude 
are large at the same wave vector, $\mib{q}=\mib {Q}$,
as shown in Fig.\ 1(a). 
Furthermore, in the real-space picture, we can understand the stability of 
the $f$-wave triplet state as follows.  
The effect of $V$ repels electrons from the nearest-neighbor sites.  
Then it is natural that the amplitude of the order parameter in real
space becomes large at the six next-nearest-neighbor sites.  
After Fourier transformation, this real-space order parameter 
results in the $f$-wave symmetry in $k$-space.  

Next we consider the electron doped case, $(n=1.2)$ (Fig.\ 3(b)). 
In this case, the obtained phase diagram differs from the 
hole-doped case.  
This difference mainly comes from the momentum dependence of 
$\chi_s$ which has peaks away from $\mib{Q}$ in contrast to 
the hole-doped case.  
As a consequence, for the singlet pairing, both spin and charge 
fluctuations mediate $d$-wave superconductivity as shown 
in Fig.\ 3(b). 
As for the triplet case, the effect of $V$ is necessary for
the stability of the $f$-wave pairing.  
The peaks in $V^t$ in Fig.\ 2(b) for $V=0$ case do not give $f$-wave 
state because the peak positions are away from $\mib{Q}$.  
Near the CDW instability, $V^t$ becomes large but $V^s$ is also 
enhanced.  As a result, $d$- and $f$-wave states compete with each
other. 

\begin{figure}
\begin{center}
\includegraphics[width=6cm]{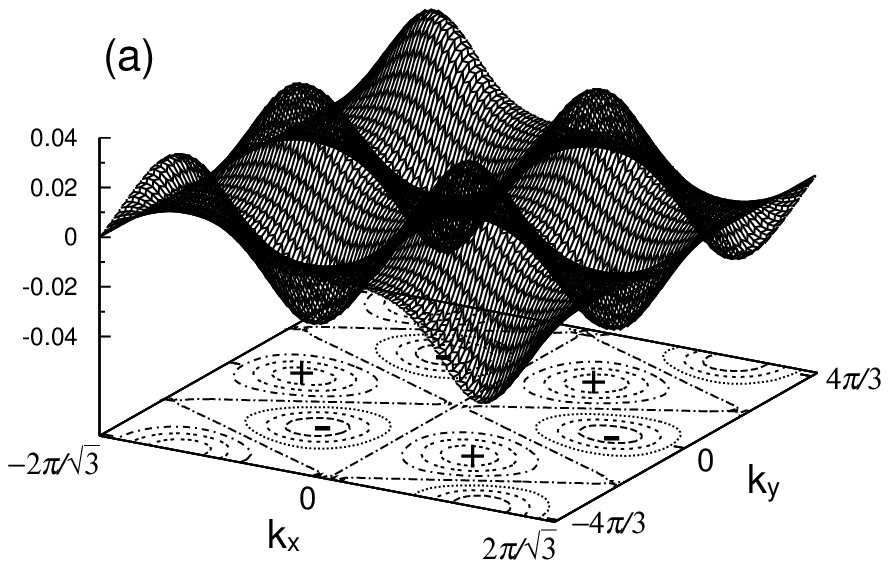}
\includegraphics[width=6cm]{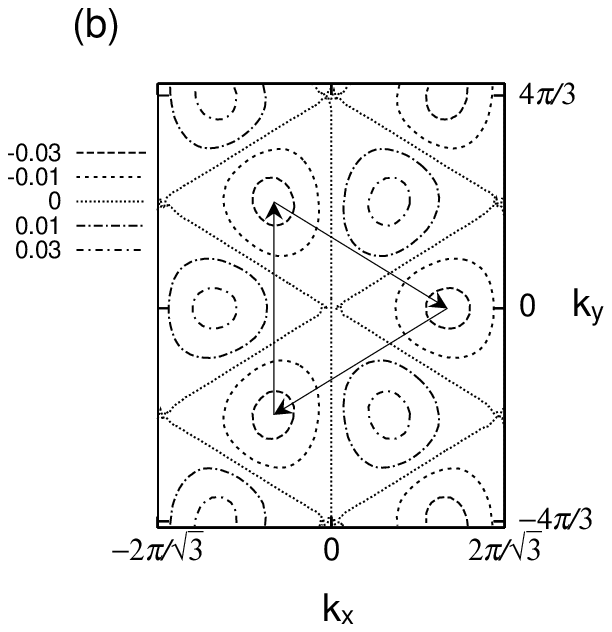}
\end{center}
\vspace{-10mm}
\caption[]{(a)The momentum dependence of the gap function
 $\Delta({\mathbf k})$ for the $f$-wave triplet pairing 
obtained at $(U,V)=(3.21,1.2)$ and $T=0.01$. The contour
plot is also shown in (b).}
\end{figure}

Experimentally it has been proposed that the triplet superconductivity 
is realized in Na$_x$CoO$_2$$\cdot y$H$_2$O from the Knight shift 
measurements,\cite{Waki} 
although there is still a controversy about the results.\cite{Nagoya} 
Furthremore the relaxation rate $1/T_1$ measurements 
in Co-NQR\cite{Fujimoto,Ishida} 
suggests the presence of the line nodes.  From these experiments 
the $f$-wave state found in this paper is promising.  

There are other possibilities for the superconducting mechanism.  
One is due to ferromagnetic spin-fluctuation.  
Although the LDA calculation shows that the Co oxides has 
$t<0$ and $n>1$, Kumar and Shastry\cite{Shastry} have proposed a model with 
opposite sign of $t$ ($t>0$), which corresponds to $t<0$ and $n<1$
in the present notation.  
Actually the high-temperature expansion study in the $t$-$J$ model 
on the triangular lattice has shown that there is a ferromagnetic 
instability in the case of $t<0$.\cite{Koretsune}
It is possible that the 
triplet superconductivity is induced by using the ferromagnetic 
spin-fluctuation in the vicinity of the ferromagnetic region.\cite{Koretsune2}  
Nearly ferromagnetic spin fluctuation has been observed 
in Co-NQR experiments.\cite{Ishida}  
The other mechanism can be in the weak-coupling theory.  
Starting from the Hubbard model on the triangular lattice or 
in a tight-binding model, $f$-wave superconductivity has been discussed 
using the third-order perturbative calculations.\cite{Ikeda}  
In order to determine which mechanism is probable, it is necessary 
to investigate the charge fluctuation in the material experimentally.  

In summary we have shown that $f$-wave triplet superconductivity 
is realized in the vicinity of CDW instability in the triangular 
lattice.  Superconductivity induced by charge fluctuation is 
a very new and interesting phenomena.  
In the square or cubic lattice, $d_{xy}$-wave singlet superconductivity 
was discussed so far.\cite{Scalapino,McKenzie,Kobayashi}
However the situation is drastically different in the triangular lattice
as shown here.

The reason why the $f$-wave state is stable is summarized as follows.
1) Charge fluctuation is equally helps singlet and triplet pairing, 
since it is the charge degrees of freedom.  
Symmetry of superconductivity is determined by the geometry of the 
Fermi surface.  
2) Due to the effect of the nearest-neighbor repulsion $V$, the 
Cooper pairs tend to be formed on the next-nearest-neighbor sites 
avoiding the nearest-neighbor sites.  
3) Since there are six next-nearest-neighbor sites in the triangular 
lattice, the sign of the Cooper pairing can take  $(+-+-+-)$ which 
fits very well to the lattice structure.  

The authors would like to thank T.\ M.\ Rice, 
P.\ A.\ Lee, B.\ S.\ Shastry, G.\ Baskaran, K.\ Ishida, K.\ Yoshimura, 
Y.\ Suzumura, and Y.\ Tanaka for very useful discussions.

\end{document}